\documentclass[aps,prl,showpacs,twocolumn,letterpaper,superscriptaddress,floatfix]{revtex4-1}
\usepackage{amsmath, amssymb,bm}
\usepackage{natbib}
\usepackage{graphicx}
\usepackage{epsfig}
\usepackage{dcolumn}
\newcolumntype{d}[1]{D{.}{.}{#1}}
\usepackage{array}
\usepackage{tabularx}
\newcolumntype{L}[1]{>{\hsize=#1\hsize\raggedright\arraybackslash}X}
\newcolumntype{R}[1]{>{\hsize=#1\hsize\raggedleft\arraybackslash}X}
\newcolumntype{C}[1]{>{\hsize=#1\hsize\centering\arraybackslash}X}

\newcommand{\be}{\begin{equation}} 
\newcommand{\ee}{\end{equation}}
\newcommand{\ba}{\begin{eqnarray}} 
\newcommand{\ea}{\end{eqnarray}}

\newcommand{\ssec}[1]{\emph{#1}.---}

\begin{document}

%Title of paper
\title{Quantum Monte Carlo calculations of neutron matter with non-local chiral interactions}

\author{Alessandro Roggero}
\email[]{roggero@science.unitn.it}
\affiliation{Physics Department, University of Trento, via Sommarive 14, I-38123 Trento, Italy}
\affiliation{INFN-TIFPA, Trento Institute for Fundamental Physics and Applications}
%\homepage[]{Your web page}
%\thanks{}
%\altaffiliation{}

\author{Abhishek Mukherjee}
\email[]{mukherjee@ectstar.eu}
\affiliation{ECT*, Villa Tambosi, I-38123 Villazzano (Trento), Italy}
\affiliation{INFN-TIFPA, Trento Institute for Fundamental Physics and Applications}
%\homepage[]{Your web page}
%\thanks{}
%\altaffiliation{}

\author{Francesco Pederiva}
\email[]{pederiva@science.unitn.it}
\affiliation{Physics Department, University of Trento, via Sommarive 14, I-38123 Trento, Italy}
\affiliation{INFN-TIFPA, Trento Institute for Fundamental Physics and Applications}

\date{\today}

\begin{abstract}
  We present fully non-perturbative quantum Monte Carlo calculations with non-local chiral effective field theory (EFT)
  interactions for the ground state properties of neutron matter. 
  The equation of state, the nucleon chemical potentials and the momentum distribution in pure neutron matter
  up to one and a half times the nuclear saturation density are computed with a newly optimized chiral EFT interaction at next-to-next-to-leading 
  order. This work opens the way to systematic order by order benchmarking of chiral EFT interactions, and \emph{ab initio} 
  prediction of nuclear properties while respecting the symmetries of quantum chromodynamics.
\end{abstract}

%\keywords{}

\maketitle
\ssec{Introduction} 
The accurate prediction of the dynamics of a supernova explosion and of the structural properties of compact stars is tightly related to the correct understanding of the properties of dense matter, and in particular of its equation of state (EoS). The conditions of temperature and density in the core of a neutron star are such that a perturbative approach based on the fundamental theory of strong interactions, i.e. quantum chromodynamics (QCD) is not possible. Non perturbative calculations could be carried on in the framework of lattice QCD.
However, at present, fully non-perturbative lattice QCD calculations of many nucleon are unfeasible,
although the extraction of nuclear forces from lattice QCD is a topic of intensive current research,
and impressive progress has been made in recent times \cite{Aoki2011, *Savage2012}.

An alternate bridge from QCD to low energy nuclear physics is provided by the use of nucleons as basic non-relativistic degrees of freedom, determining their mutual interactions by means of the chiral effective field theory (EFT).
Chiral EFT gives a systematic expansion for the nuclear forces at low energies based on the symmetries and the symmetry breakings of
QCD \cite{Epelbaum2009, *Machleidt2011, *Hammer2013}. Chiral interactions have already been employed in calculations of nuclear structure and reactions
of light and medium-mass nuclei \cite{Kalantar2012, *Barrett2013, *Roth2011, *Roth2012,*Hergert2013, *Epelbaum2009A, *Epelbaum2010, *Epelbaum2011,*Epelbaum2013, *Hagen2012A, *Hagen2012B, *Otsuka2010, *Holt2012, *Holt2013a, *Soma2013, *Wienholtz2013}, and nucleonic matter \cite{Kaiser2002, *Holt2013b,*Tews2013,*Baardsen2013,Hagen2014, Gezerlis2013, Kruger2013}.

For any given Hamiltonian, quantum Monte Carlo (QMC) methods have proven to be most accurate 
for computing ground state properties \cite{Kalos2009book,*Nightingale1998book,*Ceperley1995,*Anderson2007,*Foulkes2001,*Binder1995}. Some of the most accurate calculations for light nuclei and neutron matter
were indeed performed using continuum diffusion based QMC methods \cite{Pudliner1997,*Pieper2008, *Gandolfi2009,*Gezerlis2010}, in conjunction with the semi phenomenological 
\emph{local} Argonne-Urbana family of nuclear forces \cite{Wiringa1995,*Pieper2001}.

In general, the interactions obtained from chiral EFT are \emph{non-local}, i.e., explicitly dependent on the relative momenta of the particles. 
It is difficult to incorporate non-local interactions in standard continuum QMC methods.
Recently, an interesting approach was proposed in Ref.~\onlinecite{Gezerlis2013}, where all the non-localities up to  next-to-next-to-leading order (NNLO) 
were traded for additional spin-isopin operator dependence. This local chiral NNLO interaction was then included in a conventional auxiliary-field 
diffusion Monte Carlo (AFDMC) calculation. In this scheme, the residual non-localities would have to be treated perturbatively (See also, Ref.~\onlinecite{Lynn2012}).

In this letter, we introduce an alternative and complementary approach, viz.
performing fully non-perturbative QMC calculations with  the full non-local chiral interactions with the help of the newly developed  configuration interaction Monte Carlo (CIMC)
 method introduced in Ref. \cite{Mukherjee2013a,Roggero2013}.
The CIMC method is similar to continuum QMC, in that the ground state wave function is
obtained by applying the power method stochastically with the help of a random walk
in the space of relevant configurations. However, in contrast to continuum QMC, in CIMC the random
walk in performed in Fock space, i.e., in the occupation number basis.
As a result, non-local interactions can be easily incorporated in CIMC. 

In this letter, we report extensive calculations with a non-local chiral interaction
in which a proper QMC algorithm is used.
We use the recently developed chiral NNLO$_{\rm opt}$ interaction \cite{Ekstrom2013},
The scattering phase shifts obtained from this interaction fit the experimental database \cite{Stoks1993} at $\chi^2 \sim 1$ for laboratory energies less than $125$ MeV.
However, the contribution from the three nucleon forces is smaller with this parametrization
than with the previous ones. 

We calculate the equation of state (EoS) and the nucleon chemical potentials in pure neutron matter  up to one and a half times the nuclear saturation density.
In addition, we also present unbiased  QMC estimates of the momentum distribution.

\ssec{Method} In CIMC, the ground state (GS) wave function, $\Psi_{\rm GS}$, is filtered out by
repeatedly applying the propagator $\mathcal{P}=e^{-\tau(H-E_T)}$ on an initial state,
$\Psi_{\rm I}$, with a non-zero overlap with $|\Psi_{\rm GS}\rangle$
\be
|\Psi_{\rm GS} \rangle = \lim_{N_{\tau} \to \infty} \mathcal{P}^{N_{\tau}} |\Psi_{\rm I} \rangle . 
\ee
Here, $H$ is the Hamiltonian, $E_T$ is an energy shift used to keep the norm of the wave
function approximately constant, and $\tau$ is a finite step in `imaginary' time $\tau=it$.

This process is carried out stochastically in a many-body Hilbert space that is spanned by all Slater determinants that can constructed from  a finite set of single particle (sp) basis states.
In this work, we use the eigenstates of momentum and the $z$ components of spin and isospin as the sp
basis. The calculations are performed in a box containing $A$ nucleons of size $L^3=A/\rho$ with periodic boundary conditions.
The size of the box $L$ is fixed by the density $\rho$ of the system.
The finite size of the box requires the sp states to be restricted on a lattice in momentum space with a lattice constant $l=2\pi/L$. 
A finite sp basis is chosen by imposing a ``basis cutoff" $k_{\rm max}$, so that
only those sp states with $\mathbf{k}^2 \leq k_{\rm max}^2$ are included. 
A sequence of calculations with increasingly large values of $k_{\rm max}$ are  performed till convergence is reached.

Sampling of new states can be performed under the condition that the matrix elements of the
propagator, $\mathcal{P}$, are always positive semi-definite.
For fermions interacting with a realistic potential this condition is never fullfilled.
This gives rise to the so-called sign-problem, which is usually circumvented  
by using a guiding wave function to constrain the random walk to a subsector of the full many-body Hilbert space in which the sampling procedure is well defined. 
This restriction of the random walk introduces an approximation which is similar
to the fixed-node/fixed-phase approximation commonly used in continuum QMC.
As explained in Refs.~\onlinecite{Mukherjee2013a} and  \onlinecite{Roggero2013}, we use the coupled cluster double (CCD) type wave functions 
as the guiding wave functions. As a result, the CIMC method provides an interesting synthesis of QMC methods and CC theory. 

We were able to extend the CIMC method to the case of complex hermitian Hamiltonians, in a way 
that preserves all the favorable properties of CIMC viz., (i) the ground state energy estimate is a rigorous 
\emph{upper bound} on the true ground state energy; (ii) this upper bound is tighter than that provided by the guiding wave function;
 (iii) there is no bias due to finite (imaginary) time step, $\tau$. Note 
that, none of the above properties (i-iii) hold for the nuclear GFMC or the AFDMC methods. Details of this generalization will be provided elsewhere. 

\begin{figure}[htbp]
  \includegraphics[width=\columnwidth]{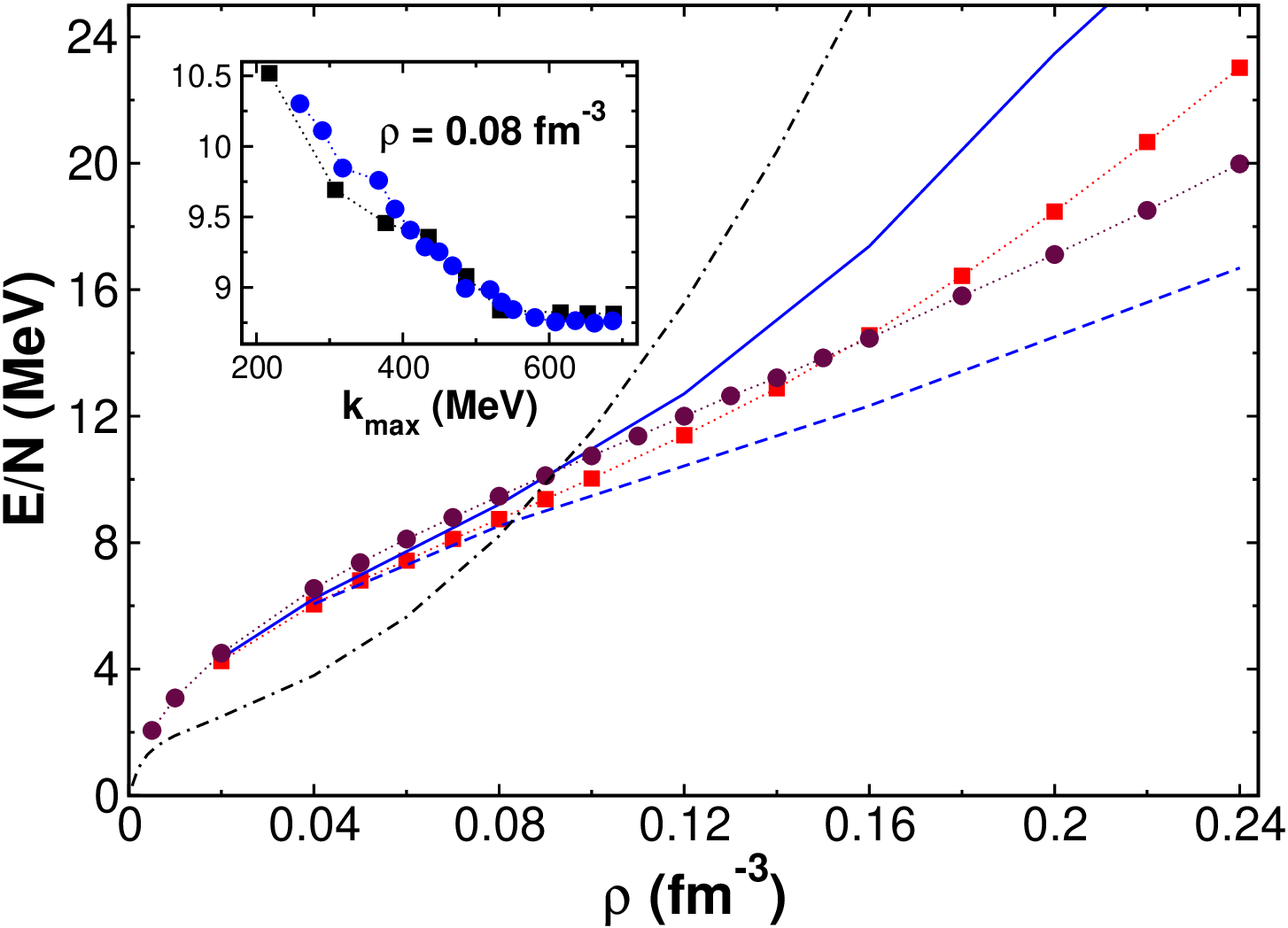}
  \caption{(Color online) The EoS of pure neutron matter: red squares - our results (66 neutrons), brown circles - AFDMC EoS with the 2b AV8' \cite{Gandolfi2013},  
  blue dashed line - APR EoS with the 2b AV18 \cite{Akmal1998}, blue solid line - APR EoS with the 2b AV18 + 3b UIX \cite{Akmal1998},  
  black dashed dotted line - NL3 EoS \cite{Shen2011}.
The inset shows the convergence of our energies as a function of $k_{\rm max}$ at $\rho = 0.08$ fm$^{-3}$ for 14 (black squares) and 66 (blue circles) particles.
The dotted lines are a guide to the eye.
\label{fig:EoS}}
\end{figure}

\ssec{Equation of state and chemical potentials}
 In Fig.~\ref{fig:EoS}, we show our results for the EoS (energy per particle vs density) 
of pure neutron matter. Energies refer to a box containing
66 neutrons with periodic boundary conditions. For periodic boundary conditions,
finite size (shell) effects are minimal for the shell closures at 14 and 66 (see, e.g. Ref.~\onlinecite{Lin2001}).
For comparison, we have also included the variational APR EoSs (two body - AV18 and two plus three body - AV18+UIX interactions) \cite{Akmal1998}, the 
AFDMC EoS (two body - AV8$^\prime$ interaction) \cite{Gandolfi2013}, and the NL3 EoS \cite{Shen2011}.

As mentioned earlier, in CIMC, successive calculations with larger sp basis sizes need to be performed till convergence. 
In the inset of Fig.~\ref{fig:EoS} we plot the energy per particle as a function of $k_{\rm max}$ at $\rho = 0.08$ fm$^{-3}$ for 14 and 66 particles. 
We deem the CIMC calculations for have converged when the difference in the energy estimate between 
successive values of $k_{\rm max}$ is less than the statistical error (typically $\sim 10 - 25$ KeV at convergence).
For all the densities considered in this work we observe a smooth convergence in the CIMC calculations as a function of $k_{\rm max}$.

\begin{figure}[htbp]
  \includegraphics[width=\columnwidth]{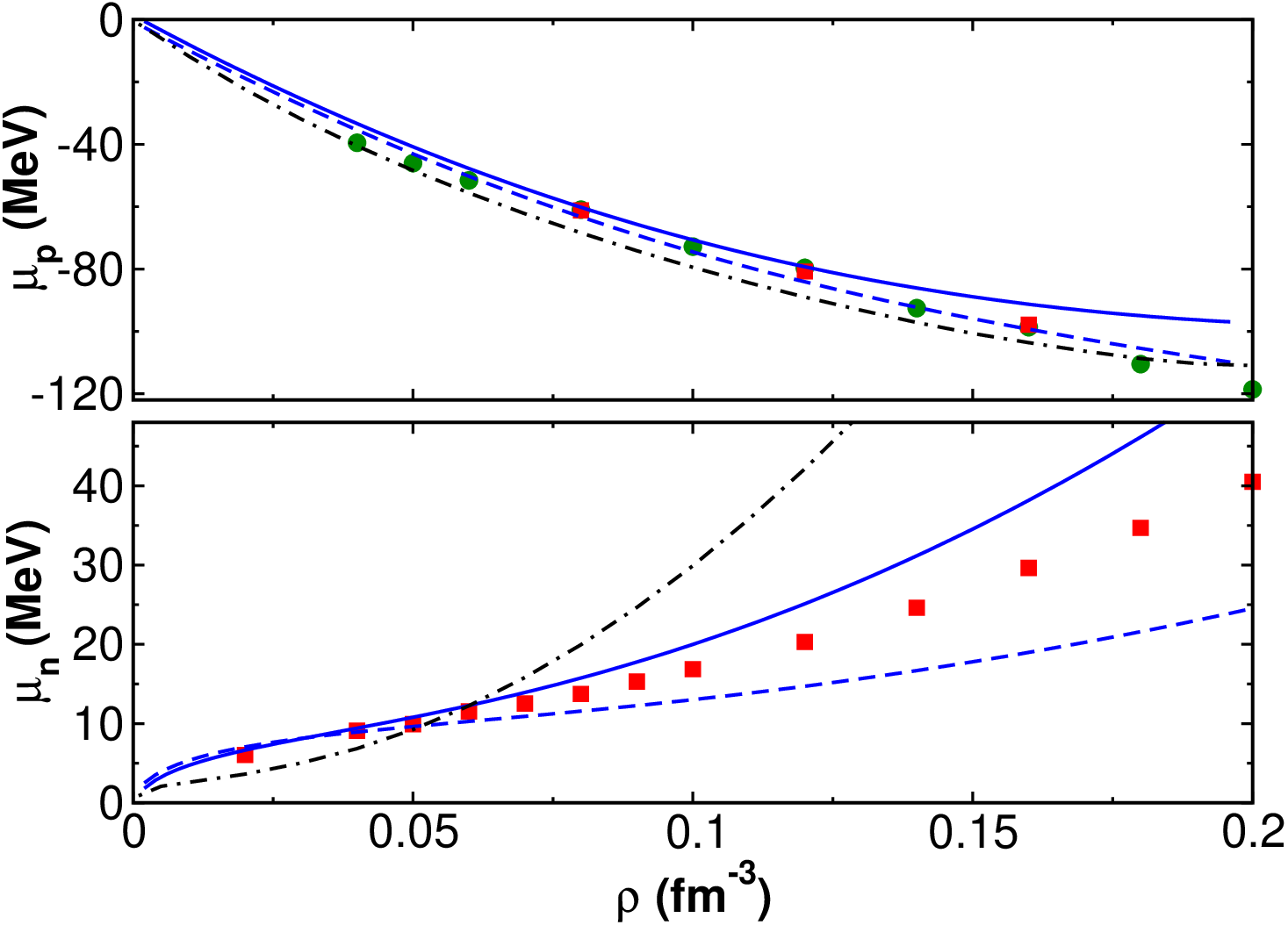}
  \caption{(Color online) The proton and neutron chemical potentials in pure neutron matter: green circles - our results for 14 neutrons. The
    rest of the legend is similar to Fig.~\ref{fig:EoS}. 
\label{fig:chem}}
\end{figure}

The nucleon chemical potentials in dense matter play a crucial role in determining
the proton fraction at beta equilibrium, and consequently the equation of state and the
cooling mechanism in neutron stars.
In Fig.~\ref{fig:chem}, we show the proton and the neutron chemical potentials
in pure neutron matter. 
We calculate the neutron chemical potential ($\mu_n = \rho \partial(E/N)/\partial \rho + E/N$) by numerical differention of the EoS.
The proton chemical potential ($\mu_p$) is calculated from the binding energy of one extra
proton in pure neutron matter. 
The calculations for $\mu_p$ were performed for 14 neutrons + 1 proton; however, we also checked
in a few cases that the results for the 66 neutrons + 1 proton case are within $2 \%$.

Most computer simulations of supernovae use phenomenological EoSs based typically on the liquid drop model, the most popular being  the Lattimer-Swesty EoS \cite{Lattimer1991}, or
on relativistic mean field theory \cite{Shen2011a, Shen2011b, *Steiner2013, *Shen2011c, *Typel2010}. As a prototype of such an EoS we have included the results from the NL3 EoS \cite{Shen2011a} in Figs.~\ref{fig:EoS} and \ref{fig:chem}. 

For $\mu_p$ all the calculations are reasonably consistent with each other. 
For the EoS and $\mu_n$, however, only the calculations based on microscopic Hamiltonians 
fit to the scattering phase shifts are consistent (within $\sim 10\%$) at low
densities ($\rho \lesssim 0.1$ fm$^{-3}$).
Other many body calculations based on microscopic Hamiltonians \cite{Hagen2014,Gezerlis2013,Tews2013,Kruger2013,Holt2013b,Kaiser2002} are also
consistent with the ones shown in the figure in this density range. 

The NL3 model on the other hand has a completely different behavior for low density neutron matter. Such a failure of most of the currently popular phenomenological EoSs  to meet the constraints set by microscopic calculations was also pointed out recently, in 
the context of chiral EFT interactions, in Ref~\onlinecite{Kruger2013}. 

\ssec{Momentum Distribution.}
 In interacting fermionic systems the momentum distribution, $n(k)$, is modified from the ideal Fermi-Dirac distribution due to quantum correlations. 
In particular, the quasiparticle renormalization factor $Z = n(k_F^-) - n(k_F^+)$
plays a fundamental role in Fermi liquid theory in quantifying the impact of the 
in-medium effective interactions \cite{Nozieres1999book}. 
In homogeneous systems, the Fourier transform of $n(k)$ is the reduced off diagonal single particle
density matrix, which is the primary object in density-matrix functional theory \cite{Gilbert1975}.  

\begin{figure}[htbp]
  \includegraphics[width=\columnwidth]{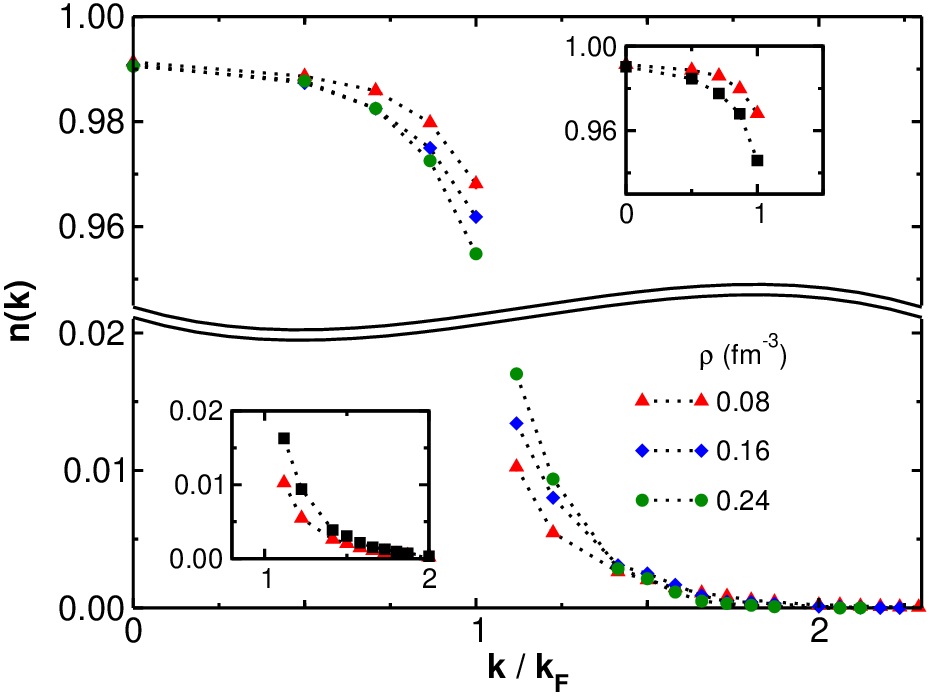}
  \caption{(Color online) Pure estimates momentum distribution of pure neutron matter from our QMC method for three different densities.
 The inset shows the comparison between the pure and the mixed estimates for $\rho=0.5\rho_0$. The dotted lines are a guide to the eye.
    \label{fig:kdist}}
\end{figure}

 In continuum QMC methods, computing an estimate of the momentum distribution independent of
 the importance function (a.k.a. pure estimator) is notoriously difficult, due to the fact that $n(k)$ is an off-diagonal operator in real space.
 In CIMC, on the the other hand, $n(k)$ is a diagonal operator. We adopt the
 method proposed in Ref.~\onlinecite{Gaudoin2007} to our CIMC method  to calculate the momentum distribution. 
In Fig.~\ref{fig:kdist} we show $n(k)$ in pure neutron matter for three different densities. 

Our estimates for the occupation number at zero momentum $n(0)$ and the renormalization factor $Z$ are given in Table~\ref{tabzn0}. 
\begin{table}[htbp]
\begin{ruledtabular}
	\begin{tabular}{ccc}
$\rho (\mbox{fm}^{-3})$   &   $Z$    & $n(0)$  \\
\hline 
\noalign{\smallskip}
0.08     & 0.9579(8)  &  0.9913(5)   \\
0.16     & 0.9484(8)  &  0.9909(5)    \\
0.24     & 0.9378(8)  &  0.9906(5)   \\
\noalign{\smallskip}
\end{tabular}
\end{ruledtabular}
\caption{The renormalization factor $Z$ and the occupation number at zero momentum $n(0)$ in neutron matter}
\label{tabzn0}
\end{table}
These results can be compared, e.g., with those in Ref.~\onlinecite{Rios2009} for CDBONN and the Argonne family of potentials
with the self-constistent Green's function method. The rather large values of $n(0)$ in Table.~\ref{tabzn0}  is due 
to the softness of the NNLO$_{\rm opt}$ interaction.

In the inset we compare the pure and mixed estimates of the momentum
 distribution at density $\rho=0.08$ fm$^{-3}$.
The mixed estimator contains an additional bias. For operators $\hat O$ other than the Hamiltonian the mixed estimator is $\langle \Psi_T|\hat O |\Psi_{GS}\rangle$, where $\Psi_T$ is the importance function used and $\Psi_{GS}$ is the ground state projected in the constrained Hilbert space. The corresponding pure estimator is instead given by  $\langle \Psi_{GS}|\hat O |\Psi_{GS}\rangle$.
We see that the biased mixed estimator for $n(k)$ overestimates both the depletion
at $k \to k_F^-$ and the growth at $k \to k_F^+$ by more than $50\%$.

\begin{figure}[htbp]
  \includegraphics[width=\columnwidth]{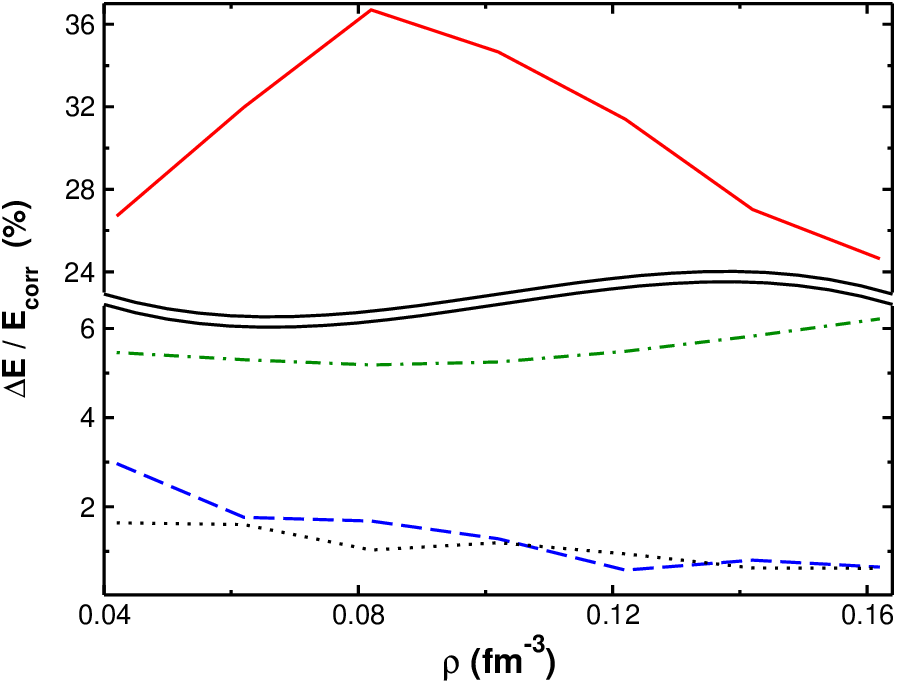}
  \caption{(Color online) Different energy scales as a fraction of the QMC correlation energy: red solid line - $\Delta E= $ difference between PT-2 and QMC energies, 
  blue dashed line - $\Delta E = $ difference between the CCD \cite{Hagen2014} and QMC energies, 
  black dotted line - $\Delta E = $ statistical error in the QMC energies,
  green dashed dotted line - $\Delta E =$ estimate of the fixed-phase bias in the QMC energies (see text).  
  \label{fig:uncert}}
\end{figure}

{\it Uncertainties of the calculations.}
 In order to make reliable \emph{ab initio} predictions, it is very important to have an
estimate of the theoretical uncertainty coming from all possible sources.
Gven the Hamiltonian and the number of particles, the uncertainties in our 
calculations come from two sources: 
(a) the inherent uncertainty in the chiral EFT Hamiltonian due to the neglect of higher orders 
and to the ultraviolet cutoff dependence, and 
(b) the uncertainty in the many body method.
In this letter we do not address the former, while noting that a significant 
amount of effort has already been devoted to this question by other authors.
The latter, in our QMC calculations (assuming convergence in $k_{\rm max}$), has two sources: the statistical error
and the bias introduced due to the fixed-phase approximation.
We see in Fig.~\ref{fig:uncert} that the statistical error is $1-2\%$ of the correlation energy (measured with respect to the Hartree-Fock energy). 
Note that this uncertainty can be systematically reduced by simply running the
simulations for longer time. For comparison, we also show 
the (absolute) difference between our QMC energy, and the energies obtained from CC theory (with the CCD wave function) \cite{Hagen2014}
and from 2nd order perturbation theory (PT-2), all as fractions of the QMC correlation energy. For this particular interaction and the 
densities considered, the CCD energy estimate is, in fact, quite close to the QMC estimate, differing at most by about $3\%$ at $\rho=0.04$ fm$^{-3}$; 
while in PT-2, the correlation energies are overestimated by $24 - 36 \%$ compared to our QMC results.

The uncertainty due to the fixed-phase approximation is very difficult to assess in continuum QMC calculations because, in general, a systematic scheme to improve the guiding wave function is not available. 
Fortunately, in our CIMC method the energies are rigorous upper bounds, and CC theory provides a systematic scheme for constructing more general guiding wave functions.
We exploit this hierarchy to provide a perturbative estimate of the leading order contribution to the 
bias due to the fixed-phase approximation, viz. that due to the exclusion of the irreducible triples in the guiding wave function.

The difference between the CCD(T) , i.e., CCD with perturbative triples, and the CCD energies \cite{Hagen2014} provide an such an estimate. However, 
just as the correlation energy is overestimated in PT-2, we expect CCD(T), which is a similar perturbative estimate, to overestimate
the residual correlation energy. 
Therefore, we obtain our improved estimate by multiplying this quantity by the ratio of our QMC correlation 
energy and the PT-2 correlation energy. Note that, the correction in energy from CCD(T) is always negative, which is consistent if one considers 
this to be the estimated correction on our QMC energy estimate,
 which is a \emph{variational upper bound}. This is not the case for the energies obtained from standard CC theory. 
 
 We plot the above estimate, again as a fraction of the QMC correlation energy in Fig.~\ref{fig:uncert}. This estimate of
  about $5-6 \%$ of the correlation energy ($\sim 1\%$ of the total energy), probably still overestimates 
 the theoretical uncertainty, since in the homogeneous electron gas, the CIMC method
 (with a CCD type guiding wave function) was found to be accurate to within $2-3 \%$ in the moderately interacting regime \cite{Roggero2013}.
In any case, the overall uncertainty in our many body method is certainly much less than the inherent uncertainty in the Hamiltonian, 
and in future work we plan to reduce it further by including the irreducible triples in our guiding wave function.

\ssec{Conclusion} In conclusion, we reported the first quantum Monte Carlo calculations with non-local chiral interactions.
Unsurprisingly, we find that the equation of state of neutron matter at low densities is reasonably model independent as long as the interaction 
used is fit to the low energy scattering phase shifts. We also provided unbiased estimates of momentum distribution and showed that the commonly used
mixed estimator grossly overestimates the depletion at the Fermi energy. 
The quantum Monte Carlo method described in this paper is quite general and can be used for nuclear matter and finite nuclei, and with three body forces.
 Work in these directions is in progress. 
\begin{acknowledgments}
We are grateful to G.~Hagen and T.~Papenbrock for sharing their CCD code and benckmarks with us,
 and to A.~Ekstr\"{o}m, G.~Hagen and T.~Papenbrock for sharing the numerical program for the chiral interaction.
 A.M. will like to thank W.~Weise for numerous fruitful discussions. 
Computations were performed at the Open Facilities at the Lawrence Livermore National Laboratory.
\end{acknowledgments}
\bibliography{chiral}

\end{document}